# Observation of X-rays generated by relativistic electrons in waveguide target mounted inside a betatron

V.V.Kaplin [1], V.V.Sohoreva [1], S.R.Uglov [1], O.F.Bulaev [2], A.A.Voronin [2], M.Piestrup [3], C.Gary [3]

[1] Federal State Scientific Institution "Nuclear Physics Institute", Tomsk, Russia,
[2] Federal State Scientific Institution "Institute of Introscopy", Tomsk, Russia,
[3] Adelphi Technology Inc.,San Carlos, USA

In this work we have observed x-ray emission from x-ray waveguide radiator excited by relativistic electrons. The experiment carried out at Tomsk betatron B-35. Such new type stratified target was mounted on goniometer head inside the betatron toroid. The target is consisted of the W-C-W layers placed on Si substrate. The photographs of the angular distributions of the radiation generated in the target by 20-33 MeV electrons have shown the waveguide effect of the three-layer structure on x-rays generated in the target. The effect proved in an angular distribution of radiation as an additional narrow peak of guided x-rays intensity inside a wide cone of usual Bremsstrahlung.

## 1. INTRODUCTION

There exist current interest in creation of the possible new types of intense x-ray sources using the relativistic electron beams passing through the periodic media and crystalline targets in order to generate x-rays on the base of transition and parametric radiations [1] or on its combinations to increase the source efficiency. Recently, the new radiators more complicated than a single crystal and a simple foil-stack were experimentally investigated. It was shown that, for example, the multicrystal radiators [2] and multilayer (x-ray mirror) radiators [3] give more intensive monochromatic x-rays than the single crystals. In general, the radiation yield is proportional to both radiator thickness and the electron-beam current transmitting through the radiator. If the radiator thickness is limited by absorption then the effect of recycling the electrons through the new radiators mounted inside a cyclic accelerator, such as, for instance, a betatron [4], is a good method to increase overall production efficiency of the new type x-ray sources. Additionally, high efficiency of slow dumping the electrons on the very narrow targets inside the cyclic accelerators allows to increase the brilliance of the x-ray sources by means of decreasing the cross-section dimension of the internal radiators. The thin-film x-ray waiveguides have been investigating [5,6] as a one from the possible ways for the production of x-ray microbeams by means of high gain compression of x-ray beam incident on the waveguides and transmitting them in the guided modes of motion. The possibility of using the multylayer structures for generation and transfer of the x-rays was investigated in [7] in the case of normal incidence of the nonrelativistic 50-100 keV electrons on a waveguide target surface. In this work we present our first attempt to use the waveguide as a x-ray source at grazing incidence of the relativistic 20-33 MeV electrons on a W-C-W three-film nanostructure created on a thin Si substrate.

## 2. EXPERIMENTAL SETUP

The experiment was performed by using the internal beam of the Tomsk betatron B-35. The scheme of the experimental setup is shown in Fig.1.

After accelerating on equilibrium orbit with radius of 245 mm, the electron beam with diameter of about 0.7 mm and divergence of about 0.001 radian was dumped on a target during 30 μc. The target was mounted on a goniometer head inside of equilibrium orbit on the radius of 210 mm. The goniometer allowed to change azimuthal position of the target and to turn the target around its vertical axis. The x-rays generated in the target went out of the toroid window closed with a 50 μm Kapton foil. The vertical and horizontal dimensions of the window were 10 and 40 mm, respectively.

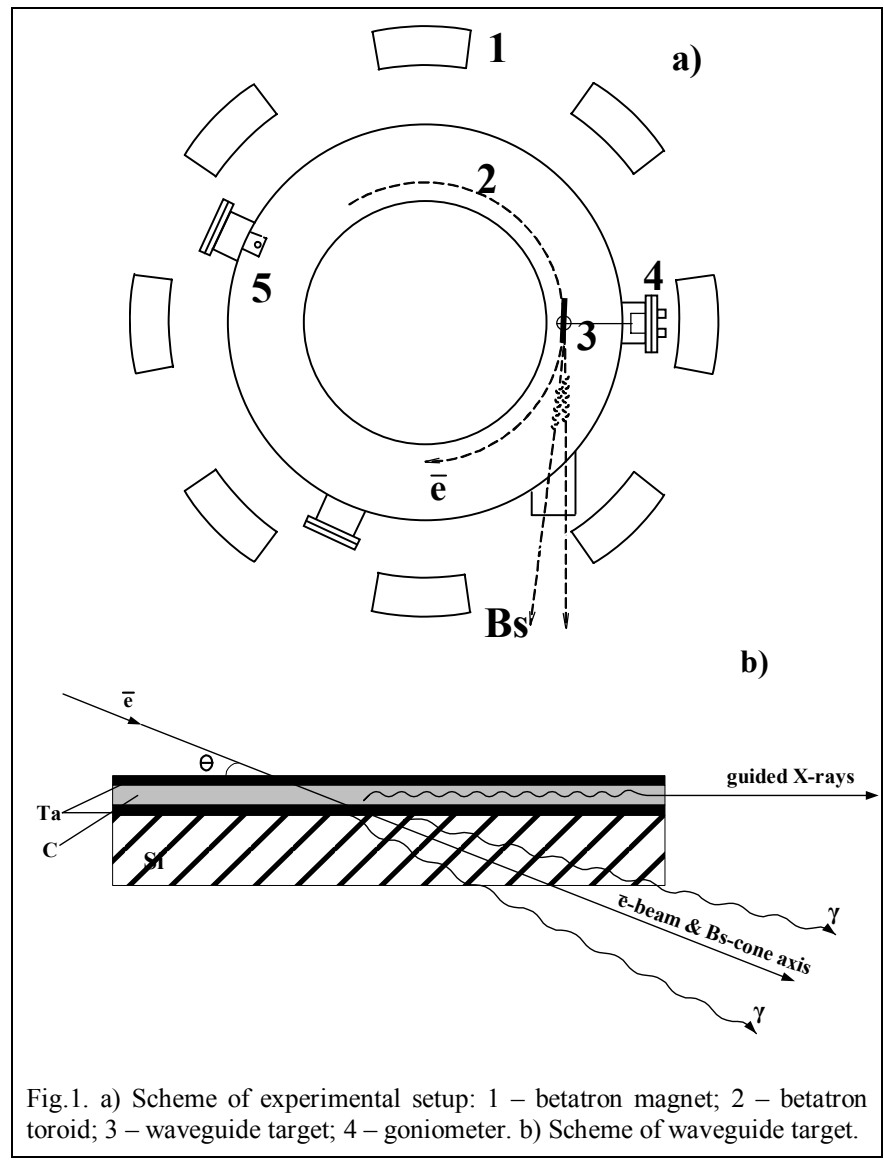

Fig.1. a) Scheme of experimental setup: 1 – betatron magnet; 2 – betatron toroid; 3 – waveguide target; 4 – goniometer. b) Scheme of waveguide target.

The cross-section of the target is schematically shown in Fig.1b. The thicknesses of the W, C layers and Si substrate were about 100 nm, 50 nm and 0.25 mm, respectively. The horizontal and vertical dimensions of the target were 2 and 10 mm, respectively. At changing the target orientation the electrons could be incident on the target surface at grazing angles in the range of 0-6 degree with respect to substrate surface or waveguide one.

The angular distributions of the x-rays generated at different target orientations and electron energies in the range of 20-33 MeV were detected with a photographic paper and films placed at the distance of 35 cm or 52 cm from the target. The photographs were then processed with a scanner to obtain the profiles of the x-ray angular distributions.

## 3. RESULTS

It was expected that at electron interaction with the target the x-rays generated in substrate and waveguide materials and/or on its boundaries might be captured in the waiveguide chanell and then transmitted the target in the guided modes of motion. As result the radiation cone must be consisted of Bremsstrahlung as a main component of the radiation generated and an additional contribution due to the guided x-rays. The width of the guided x-rays component of the radiation distribution must be much less than that of Bremsstrahlung one. It is because the waveguide modes are defined by Bragg law, and Bragg angle can not be more than critical angle θc of full reflection of x-rays from the waveguide walls. In our case of 20-33 MeV electrons the angle θc < 5 mrad for x-rays with photon energy Eγ > 10 keV is much less than the width of Bremsstrahlung cone of a few degrees.

The photographs of the x-ray beams generated at the angle θ = 0 and -30 mrad between electron beam and target surface are shown in Fig.2. The photographs demonstrate considerable difference in the shapes of the x-ray beams. The photograph (a) shows Bremsstrahlung (on the right) emitted along electron beam direction and the bright enough stripe of intensity (on the left) in target surface direction. The stripe was clearly seen when the angle between the electron beam and target surface θ was less than that of about 40-45 mrad. At θ = 0 (b), when the electron beam and target surface directions coincide with each other the Bremsstrahlung image and the stripe of intensity are seen as a single whole.

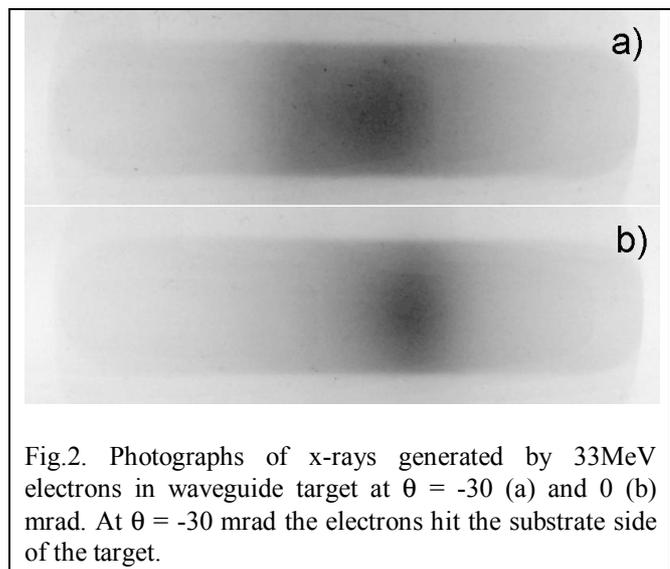

Fig.2. Photographs of x-rays generated by 33MeV electrons in waveguide target at θ = -30 (a) and 0 (b) mrad. At θ = -30 mrad the electrons hit the substrate side of the target.

The profiles of x-ray distributions along horizontal middle line of the toroid window obtained at target orientations θ = -30 and –20 mrad are shown in Fig.3. This line is perpendicular to the waveguide plane.

One can see that the x-rays cones have additional narrow peaks of highintensity in positions of the waveguide orientations. The peaks satisfy the stripes of intensity on the photographs. The narrow peak is following the waveguide orientation and increasing the intensity at decreasing the angle θ from -30 up to –20 mrad. That is to say the distribution of x-rays generated in the waveguide target is consisted of two components - a wide one defined by Bremsstrahlung and a narrow component defined by the guided x-rays. A result of the division the x-rays distribution on the principal wide and additional narrow components is shown in Fig.3. The wide component profiles are good enough discribed by the sum of the three Lorentz functions with different parameters, curves 3. The narrow component profiles, curves 1' and 2' in Fig.3, were obtained by means of subtraction of curves 3 from 1 and 2 ones, respectively. The widths of the wide and narrow components of angular distribution are about 64 and 13 mrad, respectively, and do not change at tilting the target.

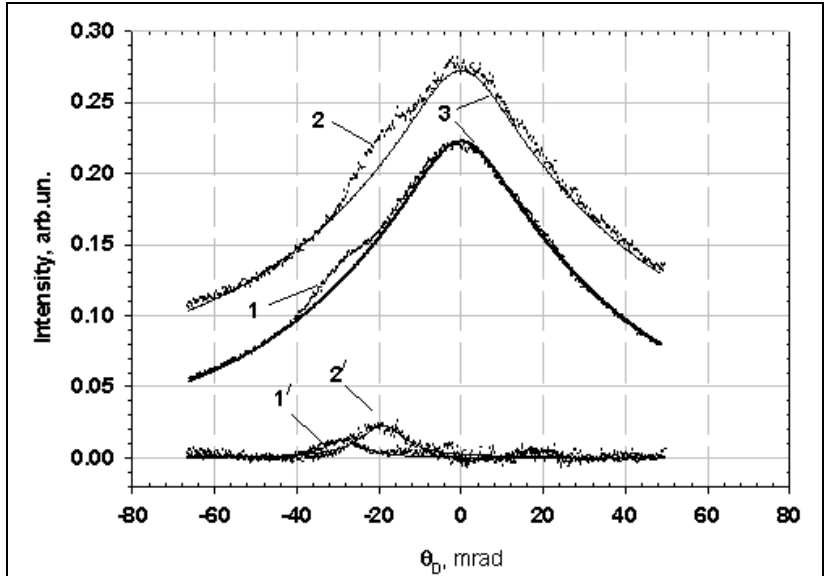

Fig.3. Profiles of cross-sections of angular distributions of x-rays generated at waveguide target orientations of –30 and -20 mrad, curves 1 and 2, respectively. Curves 3- approximation of Bremsstrahlung components of the distributions by the sum of the Lorentz functions. Curves 1' and 2' – the guided x-rays components of the distributions.

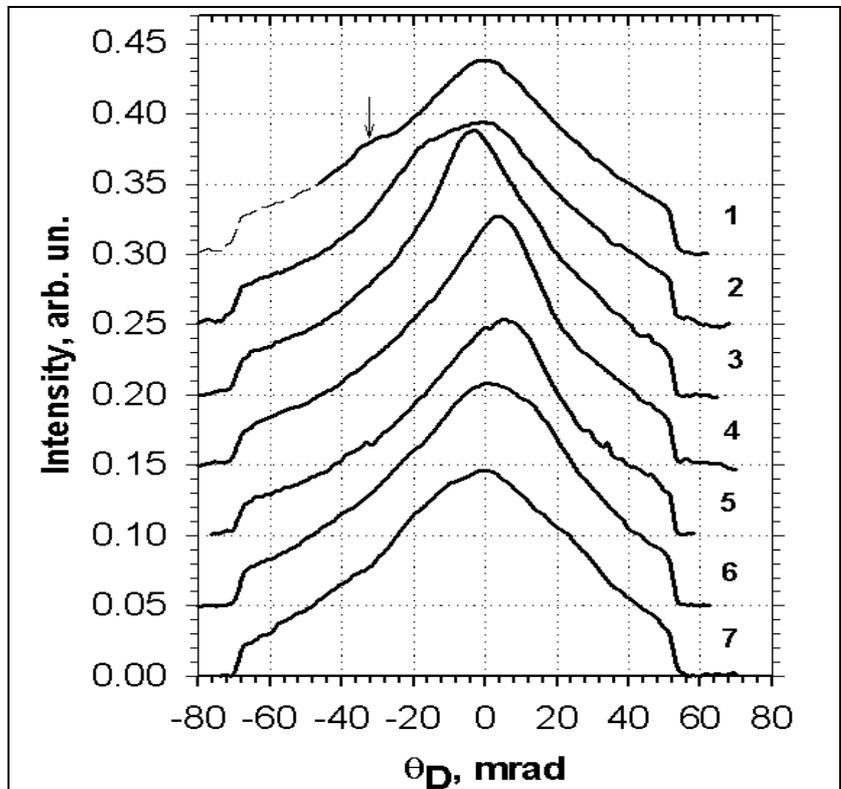

Fig.4. Profiles of the x-ray distributions measured at the angles θ = -32, -16, -4, 4, 8, 12 and 24 mrad, curves 1-7, respectively.

A set of the profiles of the x-ray distributions measured at the angles θ = -32, -16, -4, 4, 8, 12 and 24 mrad, curves 1-7, is shown in Fig.4. At decreasing the angle θ between the electron beam and waveguide directions the intensity of the guided x-rays increases, is maximal when the target is aligned along the electron beam and then decreases when the angle θ increases further. When the waveguide is aligned almost along the electron beam direction the contribution of the guided x-rays defines the shape and position of maximum of the angular distribution of the generated radiation.

The angular distributions of x-rays generated by 33 and 20 MeV electrons at waveguide orientation θ = -14 mrad are shown in Fig.5.

Comparison of the distributions is shown that at decreasing the electron energy from 33 MeV up to 20 Mev the width of the wide Bremsstrahlung component of x-ray distribution increases in about 1.5 times. But the width of the additional narrow component of x-ray distribution does not change at decreasing the electron energy. It shows again that additional contribution in the x-ray distributions is defined by the guided x-rays emitted along the waveguide direction.

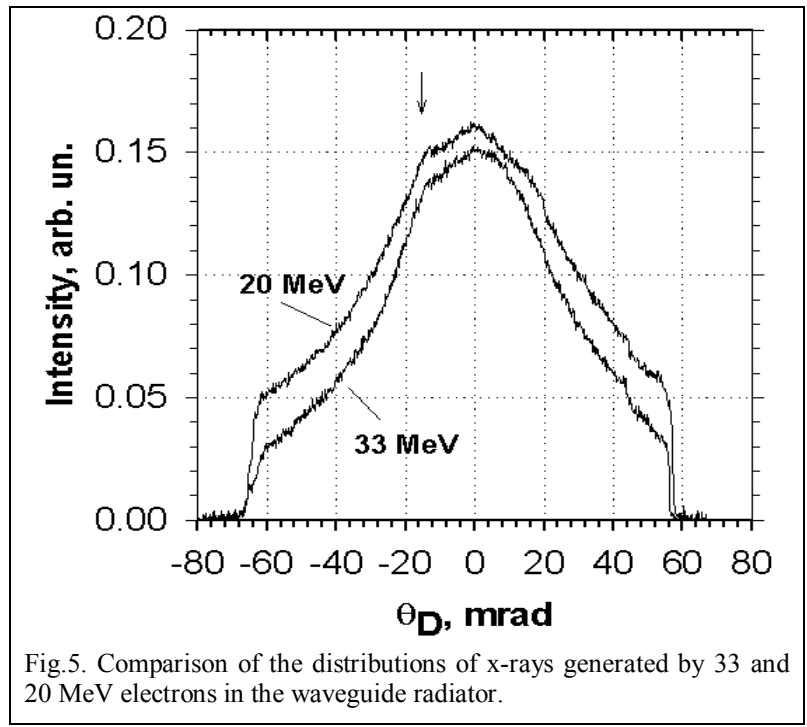

Fig.5. Comparison of the distributions of x-rays generated by 33 and 20 MeV electrons in the waveguide radiator.

## 4. CONCLUSION

Photographing the x-rays beam generated is of course as rather crude measurement of the parameters of the angular distributions. A number of moments of the effect observed must be investigated in detail with using the x-ray detectors to measure the spectral–angular characteristics of the x-rays emitted for establishing the mechanism of the guided x-rays generation. But one can already believe that observation shows clearly the existence of generation of the guided x-rays by relativistic electrons in the waveguide targets and that the effect is promising for applications.

This work is supported by the Russian Foundation for Basic Research (Project No.04-02-17-580).